\documentclass{article} 
\usepackage{latexsym}
\usepackage{graphics}
\usepackage{epsfig}
\usepackage[dvips]{color}
\usepackage[mathscr]{eucal}
\usepackage{amscd}
\usepackage{amssymb}
\usepackage{amsmath}
\usepackage{amssymb}

\newcommand{\pprec}{\prec\!\!\prec}
\newcommand{\bM}{\mathbb M}

\newcommand{\Dal}{\Box}
\newcommand{\pd}{\partial} 
\newcommand{\cO}{\mathcal O} 
\newcommand{\cN}{\mathcal N}

\newcommand{\cyl}{\mathrm{cyl}}
\newcommand{\stem}{\mathrm{stem}}

\title{Directions in Causal Set Quantum Gravity\footnote{To appear in Recent Research in Quantum Gravity, edited by A. Dasgupta (Nova Science
Publishers NY)}} 
\author{Sumati Surya \\ Raman
  Research Institute, Bangalore, India \\ \& \\ 
     Department of Physics, McGill University, Montreal, Canada \\ 
} 
\begin{document}
\maketitle 

\begin{abstract} 
  In the causal set approach to quantum gravity the spacetime continuum arises as an approximation
  to a fundamentally discrete substructure, the causal set, which is a locally finite partially
  ordered set.  The causal set paradigm was elucidated in a classic paper by Bombelli, Lee, Meyer
  and Sorkin in 1987 \cite{BLMS}. While early kinematical results already showed promise, the
  program received a substantial impetus about a decade ago with the work of Rideout and Sorkin on a
  classical stochastic growth dynamics for causal sets \cite{RS}.  Considerable progress has been
  made ever since in our understanding of causal set theory while leaving undisturbed the basic
  paradigm set out in \cite{BLMS}.  Recent highlights include a causal set expression for the
  Einstein-Hilbert action \cite{BD,Glaser} and the construction of a scalar field Feynman propagator
  on a fixed causal set \cite{stevenone,steventwo}.  The aim of the present article is to give a
  broad overview of the results in causal set theory while pointing out directions for future
  investigations.
\end{abstract}

\section{Introduction}

The origins of the causal set paradigm lie in the rich soil of Lorentzian geometry. Unlike a
Riemannian space, a Lorentzian spacetime $(M,g)$ possesses an additional structure, the {\sl causal
  structure} $(M,\prec)$ where for a pair of events $x,y \in M$, $x \prec y$ if $x$ is to the causal
past of $y$ \cite{Penrose, HE}. For a causal spacetime, i.e., one which has no closed causal curves,
$(M,\prec)$ is a partially ordered set (or poset), namely $\prec$ is
\begin{enumerate}  
\item Reflexive: $x \prec x$. \label{ref}  
\item Acylic: $x \prec y$ and $y \prec x $ implies that $y=x$. \label{acy}       
\item Transitive: $x\prec y$ and $y \prec z$ implies $x \prec z$.  \label{tran}   
\end{enumerate}       
In addition to the causal relation $\prec$ one also has the chronological relation $\pprec$ which
satisfies both acyclicity and transitivity but not reflexivity\footnote{This relation can be derived
  from $\prec$ when the spacetime is either past or future distinguishing \cite{PS}.}. Viewed in
terms of the partial order $(M,\prec)$, $M$ is merely the set of spacetime events,
stripped of its standard manifold-like topological character.

Spacetime causal structure was studied extensively in the 60's and 70's and led to a deeper
understanding of the fundamentally Lorentzian features of spacetime like black holes and their
associated singularities \cite{Penrose,HE,PK,Zeeman,GKP,HKM}. These investigations suggested that
causal structure plays a primitive, rather than a derivative role in Lorentzian geometry leading
naturally to the question -- can spacetime topology and
geometry be derived from the poset $(M,\prec)$?  

The answer to the former question has been known for several decades now, when restricted to the
class of strongly causal spacetimes \cite{Penrose, HE}. First, notice that the relational
connectivity of a generic poset ensures that it contains non-trivial topological information --
indeed, posets admit a host of different non-trivial topologies \cite{stanley,lattice}. Of these,
the Alexandroff topology $\mathcal A$ generated from a basis of Alexandroff intervals $I(p,q)\equiv
I^+(p) \cap I^-(q)$, where $I^+(p)\equiv \{s| p \pprec s \}$ and $I^-(q) \equiv \{s| s\pprec q \} $,
is of particular interest. It is not difficult to prove that in strongly causal spacetimes $\mathcal
A$ is homeomorphic to the manifold topology $\mathcal M$, or equivalently, $\mathcal A $ is
homeomorphic to $ \mathcal M$ iff $\mathcal A$ is Hausdorff \cite{PK}. It was recently demonstrated
that $\mathcal M$  can also be obtained from a more general  causal topology in  
spacetimes satisfying weaker causality restrictions \cite{PS}.  These results are quite startling
since  they tell us that the  manifold  topology,  constructed from purely
Riemannian considerations, can be derived entirely from the poset $(M, \prec)$. They provide the
first indication of the underlying primacy of the causal structure in Lorentzian geometry.

That there can, additionally, be a relation between geometry and causal structure is not altogether
surprising: it is a simple fact that the causal structure is unchanged under conformal
transformations of the spacetime geometry. However, the relationship is far more profound. Using
results due to Hawking, King and McCarthy \cite{HKM} Malament \cite{Malament,Levichev} showed that
causal bijections (i.e., maps that preserve the causal structure) between 4-dimensional future and
past distinguishing spacetimes are also conformal isometries. In other words, if a pair of
4-dimensional spacetimes admit a bijective map that preserves the causal structure then these
spacetimes are in fact conformally related.  It is possible to show that causal bijections cannot
exist between spacetimes with different dimensions ($d>2$) and hence causal structure also
determines the spacetime dimension \cite{PS}. These results imply that the conformal geometry is
completely determined by the causal space $(M,\prec)$ provided the latter satisfies certain
causality conditions. In a 4-dimensional spacetime the causal structure is therefore ``$9/10$th'' of
the spacetime geometry and the remaining ``$1/10$'' is determined by the local volume element so
that \cite{forks,reviews}
\begin{equation}\label{contmaxim} 
\mathrm{ GEOMETRY = CAUSAL\,\,\, STRUCTURE + VOLUME} 
\end{equation}   
This is one of the starting points for the causal set paradigm. 
     
The second ingredient that goes into causal set theory comes from the quantum world, and its
conjunction with gravity. This suggests that there is a fundamental Plank scale cut-off $l_p$ in
spacetime -- a suggestion that is reinforced by the apparent ``quantisation'' of the black hole area
law\footnote{ An adage attributed to Mark Kac ``Be Wise -- Discretise'' while applicable to
  various physical and mathematical situations, seems particularly appropriate to quantum
  gravity!}. How this fundamental cut-off is implemented as a ``discretisation'' of spacetime is
of course dependent on what other continuum baggage one wishes to retain. Rather than employing a
Lorentz violating cut-off like the Plank energy (or {\it length} or time), it is desirable to use
the covariantly well-defined Plank spacetime {\it volume} $V_p=l_p^4$ as a cut-off.
Introducing such  a cut-off in the poset $(M,\prec)$ produces a {\sl causal set} $C$, which apart
from being a poset is also ``locally finite''. Namely, if $J^+(x)\equiv \{y| x\prec y \} $ and
$J^-(x)\equiv \{y| y\prec x \} $, then the interval $J(x,y)=J^+(x) \cap J^-(y) $ is required to be
of finite cardinality.  The prescription of local finiteness formalises the idea of a fundamental
spacetime discreteness -- spacetime regions that have finite volume should contain only a finite
number of fundamental spacetime atoms.



Indeed, it is precisely this added ingredient of discreteness which makes it possible to 
reconstruct, in an appropriate sense, the full spacetime geometry from the causal set in the spirit
of Statement (\ref{contmaxim}). Normalising the volume with respect to a volume cut-off, this
suggests the natural correspondence
\begin{eqnarray}\label{cor} 
\mathrm{Order}  & \rightarrow & \mathrm{Causality} \nonumber \\ 
\mathrm{Number}    & \rightarrow & \mathrm{Volume}.  
\end{eqnarray}  
The particulars of course depend on the desired nature of the discrete-continuum correspondence.
Importantly, if the discrete substratum is to be thought of as fundamental, then the continuum can
arise only as an approximation.  The analogy to keep in mind is that of a fluid whose apparently
continuum nature has an underlying, discrete, molecular structure. In lattice gauge theory, or
Causal Dynamical Triangulations \cite{CDT}, in contrast, the continuum is regarded as fundamental
and is approximated by the discrete lattice. In this case, the continuum limit is essential, since
it represents the physical limit of a discrete approximating scheme. However, it is precisely
because the continuum is not a fundamental construction in the causal set approach that the
continuum {\it limit} plays no fundamental  role in the theory. As in statistical physics, one  is therefore interested in
seeing how coarse grained continuum structures emerge from the underlying causal set.

\begin{figure}[ht]
\centering \resizebox{5.0in}{!}{\includegraphics{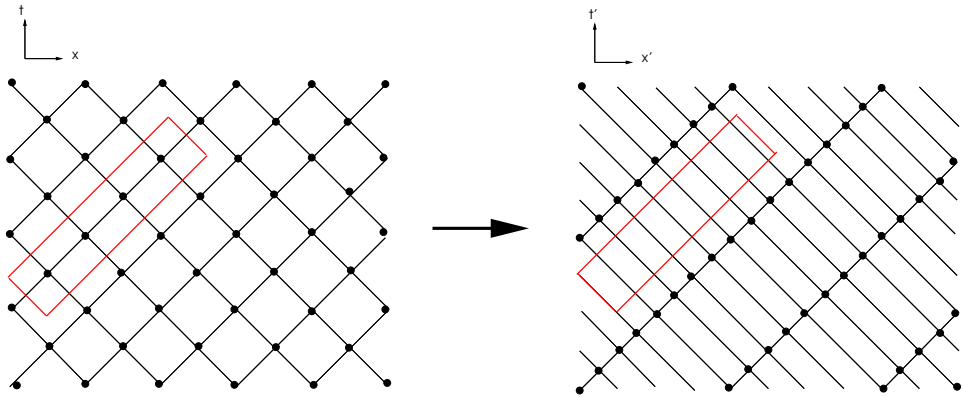}}
\vspace{0.5cm}
\caption{{\small Diamond lattice under a boost. A fixed volume region
    containing 4 elements before the boost becomes empty after the
    boost.}}
\label{Lattice.fig}
\end{figure}

\begin{figure}[ht]
\centering \resizebox{2.0in}{!}{\includegraphics{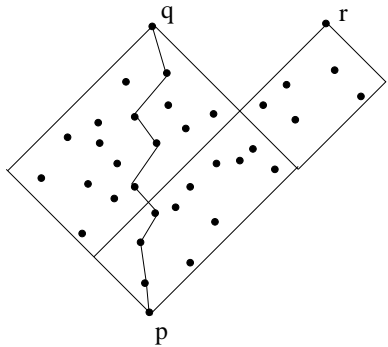}}
\vspace{0.5cm}
\caption{{\small Two Alexandroff intervals $J(p,q)$ and $J(p,r)$ in a
    causal set that is approximated by a region of 2d Minkowski
    spacetime. The length of the longest chain from $p$ to $q$ gives
    rise to the time-like distance between $p$ and $q$.  }}
\label{alexandroff.fig}
\end{figure}

However, to decide on the rules for when the approximation is a good one, it is useful to start with
the reverse question: what is the causal set that underlies a continuum spacetime? The naivest such 
discretisation of Minkowski spacetime is via a regular lattice of the sort depicted in Fig
\ref{Lattice.fig}. In the figure on the left the spacetime interval depicted has a volume $V=4V_c$
and contains precisely $4$ elements of the lattice. However, as shown in the figure on the right,
under a boost this same region contains no lattice elements\footnote{An equivalent, passive
  viewpoint, is the observation that most intervals of spacetime volume $V=4V_c$ in the figure on
  the left contain no lattice elements.}.  Thus, the second correspondence in Statement (\ref{cor})
is not consistent with a regular lattice, no matter how clever the construction.
The only discretisation that can implement the spirit of this correspondence is a ``random lattice''
of the sort described in \cite{TDLee}. Thus, a causal set $C$ can be generated from a continuum
spacetime $(M,g)$ via a random sprinkling of points ordered by the induced causal order. The most
appropriate (and possibly unique) random distribution for our purpose is the Poisson distribution
$P_V(n) = (1/n!)  \exp^{-V/V_c} (V/V_c)^n$, where $P_V(n)$ is the probability of finding $n$
elements in a spacetime region of volume $V$ for a given cut-off $V_c$.  For this distribution,
$\langle n \rangle = V/V_c$ which means that the number to volume correspondence holds on average,
thus satisfying the spirit of the Malament theorem.  Thus, one has the causal set maxim due to 
Rafael Sorkin \cite{forks}
\begin{equation} 
\mathrm{ORDER} + \mathrm{NUMBER} \approx \mathrm{SPACETIME GEOMETRY} 
\end{equation}
as a discrete analogue of (\ref{contmaxim}).  The use of the random lattice in turn has the added
bonus of preserving local Lorentz invariance \cite{BHS}. This has a profound significance for causal
set phenomenology, and one that is definitely falsifiable. As an example,  Fig \ref{alexandroff.fig} depicts a
causal set obtained by a random discretisation of a region of 2d Minkowski spacetime.

We have described how to obtain a causal set which underlies a continuum spacetime but the more
fundamental question we seek to answer is how the continuum emerges as an approximation of the
theory.  Given a causal set $C$, we will call an embedding $\Phi:C \rightarrow (M,g)$ {\sl order
  preserving} if the order relation in $C$ is mapped bijectively to the causal order induced on
$\Phi(C)$. It is an important fact that not all causal sets can be embedded via an order preserving
map into a given spacetime \cite{meyer}. In addition, the embedding $\Phi$ is {\sl faithful } if it
is order preserving and $\Phi(C) \subset M$ is a high probability Poisson distribution in $(M,g)$
\cite{luca}.  A causal set $C$ is then said to be approximated by a continuum spacetime $(M,g)$ if
$C$ can be faithfully embedded into $(M,g)$.  Note that the continuum spacetime that one obtains
via the approximation cannot be too detailed in its description --- structure (both topological and
geometrical) on scales at or smaller than the Planck scale cannot be encoded in the causal set.  It
is thus important for a faithful embedding to be approximately unique -- a given causal set should
not approximate to spacetimes that differ on scales much larger than the discreteness scale.  If
this were the case, then it means that spacetimes cannot be distinguished solely by the underlying
casual set. A Hauptvermutung or fundamental conjecture of causal set theory states \\

\noindent {\bf Hauptvermutung:} {\sl If a causal set $C$ faithfully embeds at the same density into
  two distinct spacetimes $(M_1,g_1)$ and $(M_2,g_2)$ then
  these spacetimes are related by an approximate isometry.}\\

Here, an approximate isometry captures the notion that the two spacetimes differ only at scales of
order $V_c$. The above conjecture is true when the locally finite condition is relaxed to locally
countable and dense \cite{BM}. However, the locally finite case is still unproven.  Much of
the difficulty lies in defining the closeness of two Lorentzian geometries, although progress has
been made in this direction \cite{bombelli,noldus}. On the other hand, a considerable body of
circumstantial evidence has accumulated over the years in support of the conjecture from the
construction of continuum geometrical and topological information from the causal set and we will
describe this ``kinematics'' in Section \ref{kinematics}. In Section \ref{phen} we discuss
phenomenological predictions that arise from causal sets. The most important of these is the
prediction of the value of the cosmological constant \cite{lambda}. In Section \ref{dynamics} we
discuss both the classical and quantum dynamics of causal sets and end with some concluding remarks
in Section \ref{conclusions}.

The aim of this review is to give a broad outline of causal set theory and point to directions that
are currently being pursued.  The assumption of a fundamental random discreteness means that many of
the standard tools of continuum physics cannot be used. This has been a big challenge, but one that
has been met with a fair amount of success in recent years.  After acquainting the reader with some
of these new tools, the hope of this review is to open a new window on the quantum nature
of spacetime.

\section{Kinematics}\label{kinematics}  

In this section, we will focus our attention on the question, ``Where is the geometry and topology
of a continuum spacetime $(M,g)$ encoded in a causal set $C$?'' Finding answers to this query
requires a sometimes laborious and sometimes inspired search for continuum-like structures within
the causal set.  Every successful identification of such a continuum property is a confirmation of
the Hauptvermutung since the same causal set $C$ cannot then embed into a spacetime $(M',g')$ with a
significantly different continuum property.

\begin{figure}[ht]
\centering \resizebox{2.0in}{!}{\includegraphics{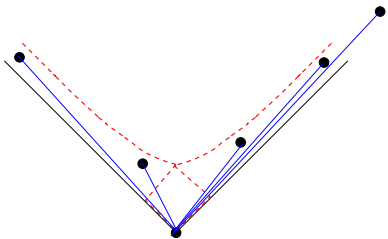}}
\vspace{0.5cm}
\caption{{\small The links from an element in a causal
    set that is approximated by flat spacetime lies all along the null
    cones. The figure shows a few (of the infinite) future
    directed links.}}
\label{valency.fig}
\end{figure}

Importantly, standard poset theory does not always provide ready answers to this question-- while a
poset itself admits several natural geometric and topological structures, these are not necessarily
appropriate in the continuum approximation.  The main difficulty arises from the randomness of the
discretisation. A useful contrast is a Regge-discretisation \cite{regge} which uses a simplicial
decomposition -- not only are topological structures retained under the discretisation, but there is
a simple procedure for calculating the local curvature using nearest neighbours. In a causal set the
nearest neighbour of an element $p$ is one that is {linked} to it: $p \prec q $ is said to be a link
if for every $ r$ with $ p\prec r \prec q$, either $r=p$ or $r=q$. A link is thus an
irreducible relation in a causal set since it cannot be deduced from transitivity. For a causal set
that is approximated by Minkowski spacetime, every element therefore has an infinite number of
nearest neighbours extending all along its future and past light-cones (see Fig \ref{valency.fig}).
This points to an inherent {\sl non-locality} in causal sets, a legacy of discreteness combined with
local Lorentz invariance \cite{rdsdice,locality}. This non-locality makes it non-trivial to
construct familiar continuum quantities and it is a happy fact that, despite this, much progress has
been made.

\begin{figure}[ht]
\centering \resizebox{2.5in}{!}{\includegraphics{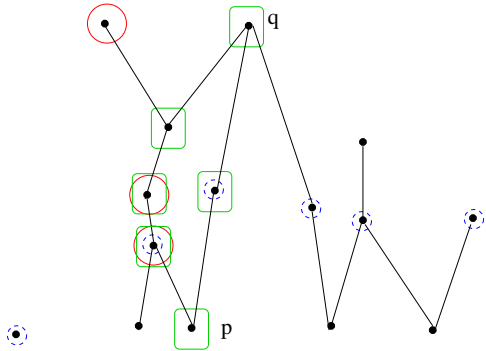}}
\vspace{0.5cm}
\caption{{\small A link, or Hasse diagram of a 14 element causal set, where only the links are
    shown. The relations can deduced from transitivity.  We show an example of a $3$-element chain
    marked by large red circles and a $5$-element antichain marked by small dashed blue circles. The
    elements in the Alexandroff interval $J(p,q)$ are marked by green boxes.}}
\label{causet.fig}
\end{figure}

The most rudimentary topological property of a spacetime is its dimension.  Let $C$ be a causal set
which faithfully embeds into a flat spacetime interval $J(p,q)$ of dimension $d$ and volume $V$.
The Myrheim-Meyer dimension estimator for this causal set uses the distribution of ``chains''. A
chain $c$ is a totally ordered subset of $C$, i.e., for every pair of elements $x,y \in c$, either
$x \prec y $ or $y \prec x$ (Fig \ref{causet.fig}).  It was shown in \cite{meyer} that for $V_c \ll
V$, the average number of $k$-element chains $\langle C_k \rangle=f(d,k) \langle N \rangle^k$, where
$\langle N \rangle = (V/V_c)$ and $f(d,k)$ is an explicit function of $d$ and $k$. This function can
then be inverted to obtain the dimension. For $k=2$, this estimator is called the Myrheim-Meyer
dimension \cite{myrheim}, where $\langle C_2 \rangle$ is simply the average number of relations in
$C$.  $f(d,2)$ is then half of the so-called ``ordering fraction'' of $C$, i.e., the ratio of the
number of relations in $C$ to the number of possible relations between $n$ elements.

Another dimension estimator is the mid-point scaling dimension, which uses the relationship between
the volume $V$ of a spacetime interval $J(p,q)$ in $\bM^d$ and the length $\tau$ of the time-like
geodesic from $p$ to $q$, i.e., $2^d= V/V'$, where the midpoint $m \in J(p,q)$ is the one that
maximises the volume $V'$ of the smaller of the two intervals $J(p,m)$ and $J(m,q)$
\cite{reviews,Gibbons}.  Numerical tests show that these dimension estimators reproduce the
continuum dimension fairly accurately for causal sets that approximate to flat spacetimes for small
$d$ \cite{meyer,reid}.

In \cite{meyer} an analysis of $\langle C_k \rangle$ for causal sets that approximate to both
anti-deSitter and deSitter spacetimes suggested that not only dimension, but also the Gaussian
curvature and the height of the conformally related flat interval could be obtained by solving a
system of three equations. The numerical results though promising, were not entirely conclusive since
relatively small causal sets were used.  In \cite{reid} a procedure was proposed for finding
Minkowski-like intervals in a class of conformally flat spacetimes from which to calculate the
spacetime dimension, and numerical simulations were carried out to verify the procedure. For large
enough causal sets these simulations work very well for the Myrheim-Meyer dimensions
\cite{reid}.

While these two dimensional estimators do seem to yield the correct values for Minkowski spacetime,
it is an important open question whether and how they can be extended to arbitrary curved
spacetimes.  It is tempting to think that these results should still be valid since dimension is
locally defined in the continuum. Indeed, in the continuum every event lies in a neighbourhood which
is diffeomorphic to flat spacetime.  Since there is no intrinsically causal definition of such
a neighbourhood, how does one construct it in the causal set?


The continuum result that the Alexandroff topology is the manifold topology in strongly causal
spacetimes suggest that this is the most natural topology for a causal set. While this will give
rise to the manifold topology in the continuum limit, what is of relevance to causal sets is not the
fine manifold topology but some coarser sub-topology. In \cite{finitary} it was suggested on general
grounds that it may suffice to consider a finitary topology generated by locally finite covering of a
space. In Riemannian spaces such locally finite sub-topologies are easy to construct from a
discretisation. For example, the circumspheres of a Delaunay triangulation of a space provide a
locally finite sub-topology.  However, Alexandroff intervals do not provide locally finite coverings
-- every element of the causal set is contained in an infinite number of such intervals in Minkowski
spacetime.

\begin{figure}[ht]
\centering \resizebox{2.5in}{!}{\includegraphics{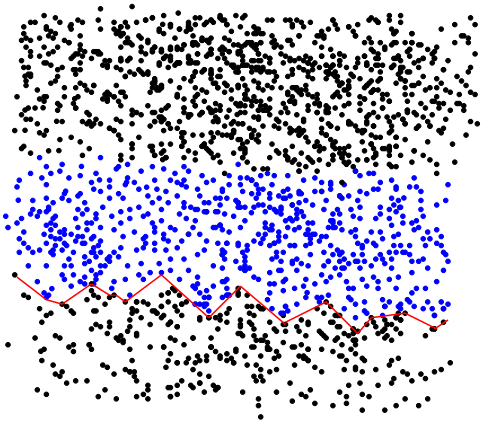}}
\vspace{0.5cm}
\caption{{\small A diagram of a causal set that is approximated
   by a region in 2d Minkowski spacetime. The
   antichain in marked as a jagged red line and the blue region denotes an
   associated  thickened antichain.}}
\label{smoothing.fig}
\end{figure}

In order to construct a locally finite covering, one can implement a localisation via the
causal set analog of a spacelike hypersurface. This is an inextendible antichain $A$ or a maximal
collection of unrelated elements in $C$ (see Fig \ref{causet.fig}). Since a choice of $A$ is a
choice of frame, one must also sample over a sufficiently large ensemble of randomly chosen
inextendible antichains.  This was the strategy adopted in \cite{hom,numhom} for constructing the
homology of a causal set which is approximated by a globally hyperbolic spacetime. Since $A$ itself
carries only the discrete topology, it needs to borrow information from the larger causal set. This
is done via a ``volume'' thickening $T_n(A) \equiv \{ x| x \succ A, |J(A,x)| \leq n \} $, where
$|.|$ denotes cardinality and $J(A,x) \equiv \{ r |\, \exists \, a \in A, \, a \prec r \prec x \}$
(Fig \ref{smoothing.fig}). The past of a maximal element $m$ of $T_n(A)$ then casts a ``shadow'' on
$A$, $J^-(m) \cap A$, and the set of shadows from each maximal element provides a covering
$\cO_v(A)$ of $A$. In \cite{hom} it was shown that the nerve simplicial complex $\cN_v(A)$
constructed from $\cO_v(A)$ gives the correct continuum homology for a large range of contiguous
values of $v$, provided $A$ satisfies certain conditions. Subsequent numerical work using ensembles of
randomly selected inextendible antichains \cite{numhom} suggests that these conditions are satisfied
quite generically.  This work lends  support to a homological version of the fundamental
conjecture, namely, that if a causal set faithfully embeds into two distinct globally hyperbolic
spacetimes, then these are ``approximately'' homological.  

Although these results are robust, they nevertheless rely on a splitting of spacetime into space and
time which is unnatural in a causal set context. It is still an open question whether interesting
topological invariants can be obtained without recourse to inextendible antichains. Since
Alexandroff intervals do not give rise to locally finite coverings of the causal set, is it
possible to get around this, and work with non-locally finite coverings? Or are there other poset
structures that one might use to define a new, more pertinent topology? Moreover, while homology is
an important characterisation of topology, there are other topological invariants, some of which
could be of importance to causal sets. Recent explorations on a Gauss-Bonnet-like relation in 2-d
causal sets are suggestive and it would be interesting to gain a deeper understanding of these results
\cite{GaussBonnet}.

The simplest geometric quantity that can be constructed in a causal set is the analog of timelike
distance. For a generic pair $x \prec y$ in $C$, there are several chains that can be constructed
with starting point $x$ and end point $y$. Each chain can be thought of as a possible time-like
curve from $x$ to $y$. Since the continuum timelike distance is obtained by maximising over the
length of timelike geodesics, a natural discrete analog of timelike distance is the length of the
longest chain between $x$ and $y$ (Fig \ref{alexandroff.fig}) \cite{myrheim}.  For a causal set that
is approximated by an interval in flat spacetime it was shown by Brightwell and Gregory \cite{BG}
that this length converges rapidly to a multiple of the proper time $\tau$, for large $\tau$. This
multiple is known for $d=1,2$ and has a fairly stringent bound for higher dimensions.  Numerical
simulations \cite{reidtwo} suggest that the timelike distance estimator holds in 2,3 and 4
conformally flat spacetimes.  More recently, progress has been made on recovering spacelike
distances in flat spacetime \cite{RW}. Extending these results to generic curved spacetime are
important but one will, again, have to address the question of localisation.

Using the antichain as a localisation device, we can also obtain the induced geometry of a spatial
hypersurface $\Sigma$, with the volume parameter defined in the homology estimation being used to
define a spatial metric. If $A$ is an antichain which is approximated by $\Sigma$ and $a_i,a_j \in
A$, one can define the predistance function $\tilde d(a_i,a_j)$ as the smallest $n$ such that
$a_i,a_j \in J^-(x)$ and $|J(A,x)|=n$ if it exists, and infinite otherwise.  This gives a labelled
graph on $A$ with vertices $a_i$ and edges labelled by $\tilde d(a_i, a_j)$. The distance function
$d(a_i,a_j)$ is then the shortest path on $A$. A similarly constructed distance function in the
continuum then allows an analytical comparison with the continuum spatial geometry \cite{geometry},
but numerical simulations have not yet been carried out.

Apart from direct topological and geometric constructions, a causal set is also characterised by the
way fields propagate on it.  The causal set analog of a Green's function for a given field is thus
an intrinsic property of that causal set.  Every causal set $C$ has a representation as a causal
matrix $M$ obtained from an order preserving labelling of $C$: $M_{ij}$ is $1$ if $i \prec j$ and
$0$ otherwise, where $i,j$ denote labelled elements of $C$. Similarly, one can construct a link
matrix $L$, for which $L_{ij}$ is $1$ or $0$ depending on whether $i$ is linked to $j$ or not.  In
\cite{alan} it was shown that for a causal set which is approximated by 2d Minkowski spacetime, 
the Green's function for a scalar field satisfies $G_{ret}+G_{adv}=\frac{1}{2}(M+M^T)$ \cite{locality,alan} and a similar
relation holds using the link matrix for causal sets that approximate to $4$-d spacetimes. The
D'Alembertian for this propagator is obtained by symmetrising $G_{ret}$ and then inverting it.
Simulations show that when restricted to causal sets that approximate to spacetime intervals, the
continuum D'Alembertian can be recovered for suitable test functions.

A different approach was adopted in \cite{locality,joeloc} for constructing the D'Alembertian
operator, which has been very successful both in resolving the problem of non-locality as well as
admitting a generalisation to curved spacetimes \cite{BD,Glaser}.  The basic insight used for taming
non-locality comes from the following observation in \cite{locality}. Consider a field $\phi(u,v)$
in Minkowski spacetime that is slowly varying in a particular frame.  If $(u,v)$ are the lightcone
coordinates in this frame, the D'Alembertian $\Dal \phi(u,v)= \pd_u \pd_v \phi(u,v)$ on an interval
$I(p,q)$ with $q=(u,v)$ and $p=(u-a,v-a)$ of area $a^2$ admits a discretisation $ \Dal_{d}
=\phi(u,v)\frac{1}{a^2}( \phi(u,v)-\phi(u-a,v)-\phi(u,v-a)+\phi(u-a,v-a))$.  Even though the system
is Lorentz invariant, the contribution from other intervals of volume $a^2$ (corresponding to boosts
of $I$) need not be the same since a given field configuration is not itself required to be Lorentz
invariant. In particular, consider the contribution from a ``narrow'' interval $I'(p',q)$ with
$p'=(u-a_1,v-a_2)$, where $a_1a_2=a^2$. As the interval becomes narrower, i.e., $a_1 \rightarrow 0$,
the first two terms and the last two terms of the D'Alembertian each sum to small values because of
the slow variation of the field. Thus, narrow intervals or those highly boosted with respect to
the slowly varying frame do not contribute significantly to the D'Alembertian. Non-locality is
embodied in a causal set by the existence of a large number of such narrow intervals of a fixed
volume and this simple observation motivated the more complicated causal set construction which we
now describe \cite{locality,joeloc}.

In a causal set the nearest neighbours are links and hence one would expect an expression for the
discrete D'Alembertian to include the sum $\sum_{y \in N_1^{-}(x)} \phi(y)$ where $N_1^-(x)$ denotes
the set of elements that are in the past of $x$ and linked to it. Since all these terms come with
equal weight, one has to counter this with terms from the next nearest neighbours, and so on. In
other words, the discrete D'Alembertian should be a sum of terms of alternating sign, with each term
of the form $\sum_{y \in N_i^-(x)} \phi(y)$, where $N_i^-(x)$ denotes the set of elements to the
past of $x$ whose order interval $J(y,x)$ with $x$, contains precisely $i-1$ elements besides $x$
and $y$.  As with any approximation, it seems reasonable to truncate the operator by including only
a finite number of layers and in order to obtain the correct continuum approximation, the exact
coefficient with which each such ``layer'' appears in the sum has to be determined.

In 2d, for a volume cut-off $l^2$, the discrete operator takes on a particularly simple form  \cite{locality}
\begin{equation}\label{nok} 
B\phi(x) = \frac{4}{l^2} (-\frac{1}{2} \phi(x) + \sum_{y \in N_1^-(x)} \phi(y) -2 \sum_{y \in
  N_2^-(x)}   \phi(y) + \sum_{y \in
  N_3^-(x)}   \phi(y) ),  
\end{equation}  
namely, there is a truncation at the third layer. The mean of this expression over Poisson
sprinklings of $C$ converges to the continuum expression for the D'Alembertian as $l\rightarrow
0$. This averaging process also allows an expression in terms of an integral kernel
\begin{equation} \label{kernel} 
B(x-y)=\frac{4}{l^4} p(\zeta) e^{-\zeta} - \frac{2}{l^2} \delta^{(2)}(x-y) 
\end{equation}  
where $p(\zeta)=1-2 \zeta + \frac{1}{2}\zeta^2$ and $\zeta l^2$ is the volume of the
order interval $J(y,x)$ in units of $l^2$. However, although the expression (\ref{nok}) 
averages to the right continuum value, the fluctuations grow like $N$. In \cite{locality} a
mesoscale $l_k >> l $ was introduced to damp these fluctuations and to rewrite the integral kernel
(\ref{kernel}) in terms of $l_k$. Discretising this kernel one gets the discrete operator 
\begin{equation} 
B_k\phi(x)= \frac{4 \epsilon}{l^2} \biggl(-\frac{1}{2}\phi(x) +
\epsilon \sum_{y \prec x} f((n(x,y),\epsilon)) \phi(x) \biggr)  
\end{equation} 
where $f(n,\epsilon)= (1-\epsilon)^n\biggl(1-\frac{2\epsilon n}{1-\epsilon}+ \frac{\epsilon^2
  n(n-1)}{(1-\epsilon)^2}\biggr) $ and $\epsilon=l^2/l_k^2$.  In writing this expression we have
reincorporated terms from more than the first three layers but in the $l_k\rightarrow l$ limit one
recovers the expression (\ref{nok}). The introduction of $l_k$ means that rather than summing over
order intervals of volumes in units of $l^2$ one is doing so over thickened layers with  volumes 
in units of $l_k^2$. This effectively damps out fluctuations although it does lead to the appearance
of a new non-locality scale. This form of the D'Alembertian has now been generalised to all
spacetime dimensions \cite{BD,Glaser,dionthesis} and a prescription for finding the exact coefficients
is given in \cite{Glaser}.

What is the generalisation of this construction to curved spacetime? It turns out surprisingly
enough that one can indeed use the very same flat spacetime expression for the operator $B_k$,
except that additional curvature dependent terms also arise \cite{BD}. Using the Gibbons-Soludukin
expression for the Ricci scalar in terms of the volume and height (or proper distance) of a small causal intervals $J(x,Y)$ \cite{Gibbons} the authors of \cite{BD} showed that 
\begin{equation} 
\lim_{l_k \rightarrow 0} B_k\phi(x) = (\Box - \frac{1}{2} R(x)) \phi(x) 
\end{equation}  
in both 2 and 4 spacetime dimensions, where $\phi$ is assumed to vary slowly with respect to $l_k$
and the radius of curvature is much larger than $l_k$. Using the constant field, this was then used
to obtain a causal set version of the Einstein-Hilbert action, the Benincasa-Dowker action in 2 and
4 dimensions
\begin{eqnarray}\label{BD} 
\frac{1}{\hbar} S^{(2)} [C]&=& N - 2 N_1 + 4N_2-2N_3 \\  
\frac{1}{\hbar} S^{(4)} [C]&=& N -  N_1 + 9N_2-16N_3+8N_4  \\  
\end{eqnarray} 
where 
$N$ is the number of elements in  the causal set $C$ and $N_i$ the number of elements in the
order interval which contains precisely $i-1$ elements besides the end points. Thus, $N_1$ counts
the number of links or 2 element chains, $N_2$ the number of 3-element chains, $N_3$ the number of
4-element chains plus the number of diamond causal sets. The Benincasa-Dowker action is a clear
proof that causal sets can give rise to local physics, and is a big milestone for the theory. The
action was generalised to arbitrary dimensions in \cite{Glaser}. 

Earlier work on finding a causal set version of the Einstein-Hilbert and matter actions for causal
sets adopted a somewhat different approach \cite{bomsver}. The Gibbons-Soludukin results were used
to find certain components of the Ricci tensor in terms of the volume and heights of small spacetime
intervals, and in order to find a way of including all possible intervals,  a minimisation
procedure was adopted. However, this procedure does not lead to a closed form for the
action. It is possible that a  procedure similar to the layer construction of \cite{locality} can be
used and this could be a useful direction to pursue.

One of the promises one tends to associate with discretisation is that of curing the UV
divergences of quantum field theory. Typical discretisations break Lorentz invariance since a UV
cut-off is put in by hand. As we have seen, causal set discretisation is very different in that it
preserves Lorentz invariance and it is therefore an open question whether it can tame these
divergences. A reasonable hope might be that causal set discretisation can  provide a Lorentz
invariant renormalisation procedure but one is currently a long way from showing this
\cite{locality}. To begin with, one has to first construct a quantum field theory on a
causal set which is approximated by Minkowski spacetime. In \cite{stevenone} a Feynman-diagram
inspired prescription was given to obtain the retarded Green's function $G(x-y)$ of a quantum scalar
field in 2 and 3 spacetime dimensions. In 2-d, one sums over all chains from $x$ to $y$, weighting
each vertex or element with an amplitude $b$ and each leg or relation that appears in the chain by
an amplitude $a$.  In 4-d one sums only over chains in which all the relations are strictly
links. Comparison of the continuum propagator with the discrete version allows the amplitudes $a$
and $b$ to be evaluated. The 2d causal set propagator is well approximated by the continuum and in
4d, this is true in the continuum limit. 

In \cite{steventwo} the next big step was taken by constructing the Feynman propagator using a
rather novel procedure. To begin with, the Pauli-Jordan function is a matrix on the causal set
$\Delta={G_R} -{G_A}$, where ${G_R}$ and ${G_A}$ are the 2d or 4d causal set versions of the
retarded and advanced Green's functions that were obtained previously. $i\Delta$ is a Hermetian
skew-symmetric matrix, and hence of even rank $2s$. This means that its non-zero eigenvalues appear
in real positive and negative pairs $(\lambda_i, -\lambda_i)$ with $i=1, \ldots, s$ . The associated
positive and negative eigenvectors $\{ u_i\}$ and $\{ v_i\}$ are used to define a matrix $Q=
\sum_{i=1}^s \lambda_i u_iu_i^\dagger$, so that $i\Delta=Q-Q^*$, and bosonic quantum field operators
$\hat\phi_x$ are associated with each causal set element.  Apart from satisfying the standard
Hermiticity and commutation conditions these operators satisfy a new condition which replaces the
standard requirement that $\hat\phi$ satisfy the Klein Gordon equation. Namely, all operators linear
in $\hat \phi$ which commute with $\hat \phi$ are required to themselves be zero. Thus 
\begin{equation} \label{newcond}
i\Delta w =0 \Rightarrow \sum_{x=1}^N w_x \hat \phi_x=0,   
\end{equation}  
where $N$ is the number of causal set elements. The positive and negative eigenfunctions $\{u_i\}$
and $\{ v_i\}$ of $i\Delta$ allows one to define creation and annihilation operators and using the
conditions on $\hat \phi_x$,  a consistent mode decomposition for the field operator is 
obtained.  The vacuum associated with the creation and annihilation operators is then used to  
construct the two point function $Q_{xy}=\langle 0 |\hat \phi_x \hat \phi_y | 0\rangle $ and the
natural order-preserving labelling on the causal set gives the requisite time-ordering. The
Feynman operator is then given by ${G_F}={G_R}+iQ$. Numerical tests in 2-d and 4-d demonstrate
that the discrete operator has the recognisable continuum form as the discreteness scale
goes to zero. 

What is makes this construction particularly interesting is that a vacuum state is obtained purely
algebraically without explicit reference to a coordinate system. The positive and negative mode
decomposition arise from properties of  the Pauli-Jordan matrix which itself is defined independent of
any frame of reference. Whether this construction has consequences for quantum field theory on
curved spacetime is clearly an important direction to pursue. What special role does the new
ingredient, Eqn (\ref{newcond}) play, for example?

The existing calculations for the blackhole area and entropy bounds can also be categorised as
kinematics since the causal set itself remains non-dynamical \cite{ddou,ddouthesis,davezohren}. In
\cite{ddou} the area of a black hole like region is calculated by counting the links that cross the
horizon. This and similar calculations require a past volume cut-off, but yield the desired
geometric result, namely that the area of the blackhole horizon is proportional to the number of
links that cross it. Whether assigning these links the term ``horizon degrees of freedom'' offers a
glimpse into a more complete description of this relationship can only be determined by a better
understanding of the quantum dynamics of causal sets.

\section{Phenomenology} \label{phen} 

An important criticism levied against quantum gravity by most other physicists is that it offers us
no experimental signatures. Almost as a response to this criticism, recent years have witnessed a
plethora of { models} of quantum gravity --- from large extra dimensions which bring down the energy
scale of quantum gravity to those in which there are violations of Lorentz invariance, one of the
most stringently verified symmetries of nature. It is therefore of interest to see what it is that
causal set theory can offer by way of an experimental signature. Without a complete theory of causal
set quantum gravity -- kinematics {\it and} dynamics -- one may worry that any model building is
futile. However, there is a big desert of energy scales between the Plank energy scale and that of
current colliders and it is a legitimate query to ask whether the causal set discretisation of a
spacetime has observable consequences. The small effects of discreteness can in principle be amplified by
cosmological distances leading to clear cosmological signatures. 

The most significant phenomenological signature from causal set theory was the prediction of a
cosmological constant in the late '80s \cite{lambda}. Indeed, this preceded the observation of a
small but non-zero value for $\Lambda$, and is of the same order of magnitude as the causal set
prediction. The basic argument is very straightforward consequence of the random discreteness for
spacetime and the number to volume correspondence (\ref{cor}). The cosmological constant appears in
the Einstein-Hilbert action as the volume term $ \Lambda \int \sqrt{-g} d^4x = \Lambda V$ where $V$
can be taken to be the total past volume of the universe. Fixing this volume in a unimodular
modification of gravity gives rise to an uncertainty relation between the fluctuations in $\Lambda$
and $\Delta$: $\Delta \Lambda \Delta V \sim 1 $ in natural units. In a causal set discretisation of
the FRW universe, the fluctuation in volume can be determined from the Poisson distribution, i.e.,
$\Delta N =\sqrt{N} $ so that $\Delta V = \sqrt{V}$.  One finds that $\rho_\Lambda \sim
\sqrt{V}^{-1} \sim H^2 = \rho_{critical}$ at all times and that given the age of our universe and
hence $V$, $\Delta \Lambda$ in our present epoch is the small value $10^{-120}$ in natural units.
Such a small value for $\Lambda$ has been notoriously difficult to calculate from standard quantum
field theory arguments and hence it is impressive that it emerges so naturally from causal set
theory. However, it is important to point out that the argument above only predicts the value for
the {\it fluctuation} in $\Lambda$. The hope is that a complete answer to this question lies within
the full causal set quantum gravity.

In \cite{ADGS} a stochastic model for the evolution of a fluctuating $\Lambda$ was incorporated into
a modified FRW model which retains only the Hamiltonian constraint. The stochastic models are
characterised by a single free parameter $\alpha$ and for suitable choices of $\alpha$ numerical
simulations show a $\Lambda$ which fluctuates between negative and positive values with a
significant fraction of the trials giving rise to a small but positive cosmological constant in the
present epoch. In these same trials, the Big Bang Nucleosynthesis phase is characterised by a large
negative $\Lambda$ which the authors suggest may help explain the reduced Helium abundance of
observations. Expanding this stochastic model to one in which $\Lambda$ varies spatially is of great
importance in getting observational constraints on the free parameters \cite{constraints}.
Moreover, an answer to why the mean value of $\Lambda$ is zero may only be possible in a more
developed cosmological model which incorporates quantum fields.

As we have noted, the basic argument for a fluctuating $\Lambda$ is very simple and seemingly
transcends the details of the theory. Indeed, it has been suggested that since spacetime
discreteness is a standard feature of several approaches to quantum gravity, they must also admit a
similar solution to the cosmological constant problem. This would in principle be reasonable if we
could view this problem purely as a macroscopic manifestation of a fundamental theory. After all,
several disparate approaches to quantum gravity yield the same black hole area law. However, it is
the precise manifestation of causal set discreteness via a number to {\it spacetime} volume
correspondence that distinguishes it from the other approaches and is key to the arguments of
\cite{lambda}. If this correspondence is taken seriously, then one is forced to consider a random
discretisation of spacetime and no other approach to quantum gravity currently uses this as a
starting point.

Are there other observational consequences of causal set discreteness? The Poisson sprinkling of
causal set elements allow for the possibility of very large voids or continuum volumes which are
entirely unpopulated by causal set elements. Such voids, if ubiquitous, would surely contradict
experimental observation since in such regions the continuum description would entirely
breakdown. In spacetimes of infinite extent it is indeed the case that there will almost surely be
arbitrarily large voids, but it is more pertinent to ask this question of our observable
universe. In \cite{swerves} it was argued that the probability of a single nuclear sized void in the
observable universe is less than $ \sim 10^{252} \times e^{-10^{72}}$! Thus, one needs to find
more significant observational signatures for causal set discreteness.

In \cite{swerves} a model of a massless classical particle propagating on a causal set was
constructed. A causal set $C$ which faithfully embeds into flat spacetime offers no straight line
path for a particle which thus must ``swerve'', i.e., the direction of the momentum must change
while hopping from one element $e_n$ to the next $e_{n+1}$. The momentum $p_{n+1}$ associated with
such a hop can be taken to be proportional to the spacetime vector in the spacetime embedding, with
starting point $e_n$ and end point $e_{n+1}$. Since this is a classical motion, the particle is
assumed to choose a path in $C$ which is as close to a straight-line as possible. This means that,
given a particular pair $(e_n,p_n)$, the next element $e_{n+1}$ is chosen as close to the direction
$p_n$ as possible so that $|p_{n+1}-p_n|$ is minimised. Additionally, $e_{n+1}$ is chosen to lie
within a certain proper time $\tau_f$ which corresponds to a ``forgetting time''. In the continuum
this process manifests itself as a proper time diffusion in phase space $\mathbb H ^3 \times \mathbb
M^4$, where $\mathbb H^3$ is the momentum mass shell and $\mathbb M^4$ is Minkowski spacetime. The
probability distribution $\rho(p^\nu, x^\mu, \tau)$ then satisfies the Lorentz invariant diffusion
equation \cite{swerves, manifoldofstates}
\begin{equation} 
  \frac{\partial }{\partial \tau} \rho = k \Delta_p^2 \rho -
  \frac{1}{mc^2}p^\mu\frac{\partial}{\partial x^\mu} \rho,   
\end{equation} 
where $k$ is a parameter which will depend on the forgetting time $\tau_f$.  Assuming the
probability distribution to be spatially uniform, the Gaussian evolution of the probability in
momentum space means high energy events become significant over sufficiently long periods of
time. This affords a Fermi-acceleration type mechanism for generating very high energy cosmic
rays. For protons, laboratory and astrophysical data can be used to put too stringent a limit on $k$
for protons \cite{kalopermattingly}.  Nevertheless, the possibility of neutrino sources giving rise
to similar signatures hasn't yet been ruled out. Since the mechanism does not depend on a specific
matter distribution, if there are causal set origins for ultra high gamma rays then these should be
distributed isotropically. Moreover, as in a standard diffusion process the swerves do not conserve
energy, but as discussed in \cite{mattingly} this does not contradict the current astrophysical
bounds on Lorentz violation.  Subsequent work in this direction includes the massive particle models
of \cite{dpsone,philipotthesis}.

More recently, a model has been proposed to study the effects of Lorentz invariant discretisations
on the CMB polarisation \cite{pol}. Here, in addition to the phase space one has the set of
polarisation states $\mathcal B$ given by the Bloch sphere. Neglecting effects of phase spacetime
diffusion, one gets a Lorentz invariant diffusion on $\mathcal B$ which reduces for linearly
polarised states to a diffusion cum a drift equation. In the cosmic frame these are frequency
dependent. The authors of \cite{pol} propose that this can give rise to signatures in CMB
polarisation and the hope is that this exciting proposal will find useful constraints when new data
becomes available from Planck. 

\begin{figure}[ht]
\centering \resizebox{2.5in}{!}{\includegraphics{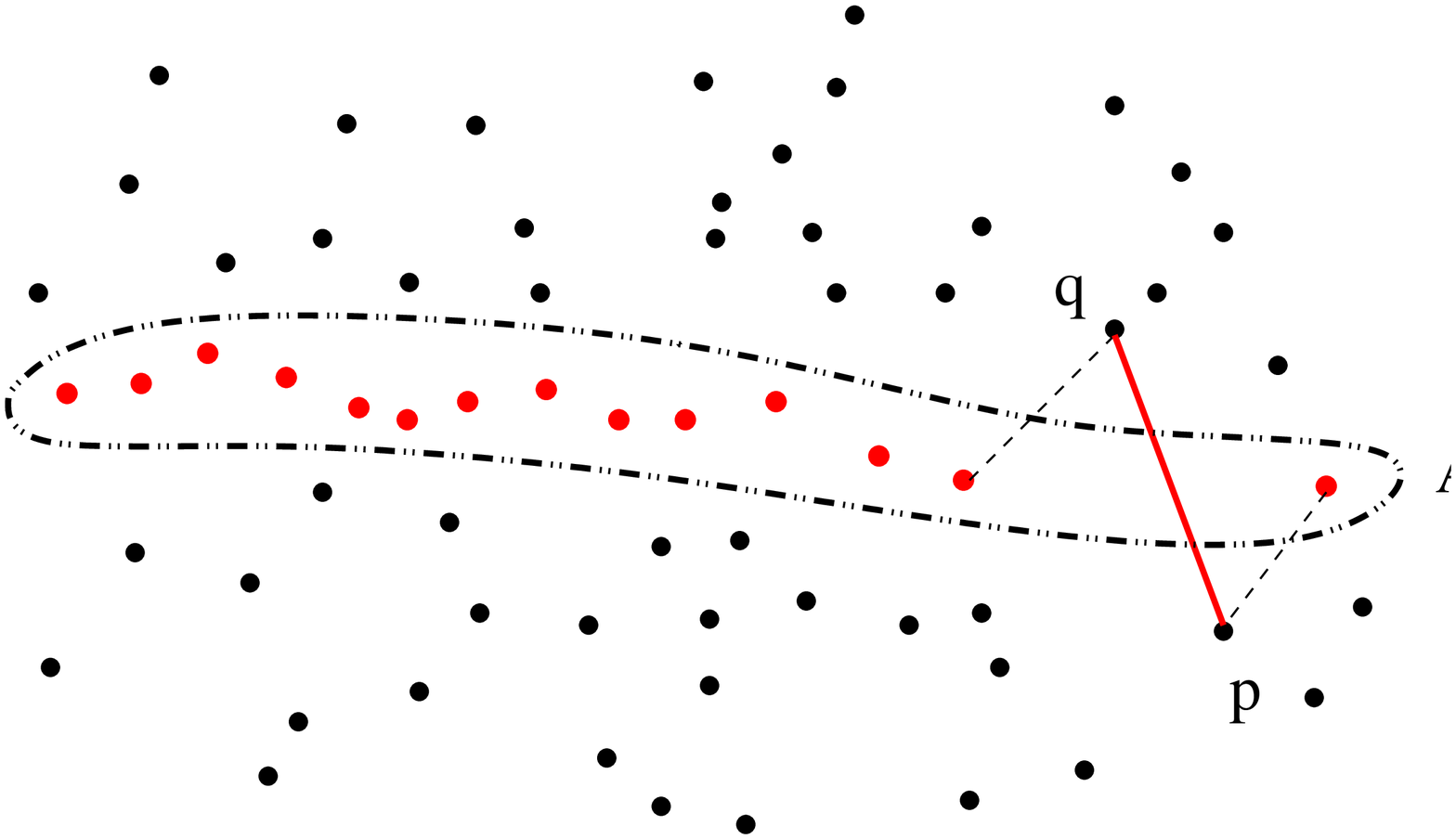}}
\vspace{0.5cm}
\caption{{\small A missing link $p \prec q $ through an inextendible
    antichain $A$ is marked with a bold red line.}}
\label{missing.fig}
\end{figure}

Before going on to examining dynamics, it is perhaps important to emphasise that while it may seem
tempting to some to construct a $3+1$ Cauchy-type formulation of causal sets based on thickened
antichains, such an attempt is doomed to failure for a reason that can, once again, be traced to
non-locality. Instead of being able to mimic a Cauchy surface whose intrinsic and extrinsic geometry
captures all the required information for future evolution, an inextendible antichain $A$ is like a
``sieve''-- a large amount of geometric information in fact by-passes it, because of what I will
call ``missed links''. Consider a causal set $C$ that faithfully embeds into Minkowski spacetime and
let $A$ be an inextendible antichain in $C$.  While $J^+(A) \cap J^-(A) =A$, where $J^+(A) \equiv \{
r| \, \exists \, a \in A, \, a \prec r \}$ and similarly $J^-(A)\equiv \{ r| \, \exists \, a \in A,
\, r \prec a \} $, so that $A$ neatly divides the {\it elements} of $C$ into the disjoint subsets
$C=A \sqcup \, ( J^+(A)\backslash A) \, \sqcup \, (J^-(A)\backslash A)$, this is not true of the
relations. In particular there exists pairs $p,q$ with $p \in J^-(A)\backslash A $ and $q \in
J^+(A)\backslash A $ such that $p \prec q$ is a link, which $A$ however ``misses'' (see Fig
\ref{missing.fig}).
In fact $A$ misses an infinite number of links in Minkowski spacetime. Are these causal links
relevant to signal propagation \cite{rdsdice,swerves,dpsone,dhstwo}? If so, then far from
capturing all the information relevant to its future, $A$ in fact lets most of it slip through! A
finite volume thickening of $A$ clearly cannot eliminate this problem, and one has to contend with
the question of how to recover the approximate global hyperbolicity of our physical laws. It is
important that the missed links tend to be those that are highly boosted with respect to the rest
frame associated with the antichain and hence it seems physically reasonable that such a signals
will tend to pass through a macroscopic body unnoticed.  However, given that there are an infinity
of such links in Minkowski spacetime it is important to know whether the effect is cumulative.  Does
it imply, for example that in order for causal sets to be compatible with known physics, that the
universe be past finite, at least until the last cosmological bounce?

\section{Dynamics} \label{dynamics}

\begin{figure}[ht]
\centering \resizebox{3.0in}{!}{\includegraphics{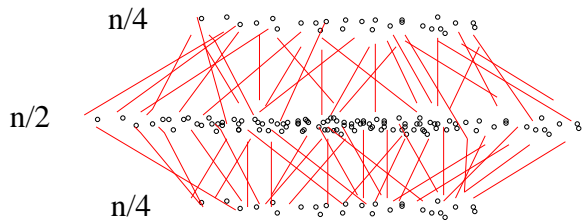}}
\vspace{0.5cm}
\caption{{\small A three layer $n$ element Kleitman-Rothschild poset}}
\label{KR.fig}
\end{figure}

Given that causal sets do not admit a natural space and time split, it is clear that a Hamiltonian
framework is not viable for describing causal set dynamics. Indeed, it is far more natural to talk
of dynamics in a path integral or sum-over histories language.  In a path integral representation of
4-dimensional quantum gravity, the histories space is the space of all Lorentzian 4-geometries. In
causal set theory this is replaced by the set of all (unlabelled) locally finite posets
$\Omega$. Importantly, $\Omega$ includes causal sets that do not have a continuum approximation, and
the hope is that a suitable dynamics would, in the classical limit, pick out causal sets that are
approximated by continuum spacetimes. But does a typical causal set look anything like a spacetime
and if not does it only differ from it ``in the small''? Dynamical triangulations have already
taught us that such an expectation is naive, however reasonable the discretisation may seem -- it is
a well know problem that most simplicial manifolds have a branched polymer structure, which dominate
the path integral \cite{polymer}. Causal sets also potentially face such an entropy problem. As the
size of the causal set $N$ gets larger, most causal sets are of the Kleitman-Rothschild or (KR) form
shown in Fig \ref{KR.fig} \cite{KR}. These have only three layers, with a large number of
relations between each layer and are therefore most unlike spacetime. However, their  numbers
grow like $2^{N^2/4}$ where $N$ is the size of the causal set and hence seem like a threatening
presence in the path integral. 
Moreover, it was shown in \cite{dhar} that apart from the KR posets there is a hierarchy of classes
of sub-dominant causal sets, with the number of layers increasing as one goes down the
hierarchy. The entropic landscape is therefore complex and characterised by first order transitions
between these sub-dominant classes \cite{dhar}.  Any causal set dynamics must therefore work hard to
counter this dense entropy landscape.

It is thus an important and surprising result that a dynamics can be constructed which counters this
entropy. These are the Rideout-Sorkin classical sequential growth models alluded to in the abstract
\cite{RS,davethesis}. These models use a bottom-up approach to dynamics -- rather than deduce a rule
from the continuum one begins with a set of fundamental principles. Starting from the empty set, a
causal set is grown element by element in a Markovian fashion by assigning probabilities to each
transition from an $n$ element causal set to an $n+1$ element causal set. The $n+1$th element can be
added either to the future of or be unrelated to an existing element, but it cannot be added to its
past.  Apart from being Markovian, the dynamics is required to obey label invariance (a discrete
avatar of covariance) as well as a causality condition dubbed Bell causality.  The latter encodes
the requirement that the transition probabilities should be independent of the ``spectator''
elements, i.e., those that are not involved explicitly in the transition.  These simple rules can be
manipulated to extract a set of coupling constants, $t_n$, one for each stage of the growth, which
then fully determine the dynamics. In the cosmological context, starting from generic initial
conditions, these coupling constants are seen to get renormalised after every cosmological epoch
separated by a ``bounce'', with a line of fixed points corresponding to a simple, transitive
percolation dynamics, with $t_n=t^n$ for $t\geq 0$ \cite{CMRS}. This last era of transitive
percolation gives rise typically to causal sets which have more than a fleeting resemblance to
continuum spacetime \cite{davemaqbool} although they do not seem to admit an exact spacetime
approximation. Nevertheless they teach us that it is possible to overcome the threat of KR posets
with an appropriate choice of dynamics.

Since the growth process cannot add a new element to the past of an existing one, the stochastic
growth models give rise to causal sets that are past finite. Each causal set is moreover labelled by
the specific growth process and hence, even though the probabilities are label invariant, the causal
sets themselves are labelled. Because of the nature of the growth the labelling is order preserving,
i.e., $x\prec y \Rightarrow l(x) < l(y)$ where $l(x) \in \mathbb N$ denotes the labelling of the
causal set element $x$.  Thus, the space of histories or sample space $\Omega$ is the space of past
finite labelled causal sets. An obvious and important question is -- what are the covariant or label
invariant questions that one can ask of such a system? This question was addressed in
\cite{observables} by re-expressing the dynamics as a probability measure space. A finite $n$-element
labelled causal set $c_n$ can be used to construct a cylinder subset $\cyl(c_n)$ of $\Omega$ which
is the set of all labelled causal sets in $\Omega$ whose first $n$ elements are the causal set
$c_n$. The event sigma algebra $\mathcal S$ is then constructed by taking countable unions and
intersections of all such cylinder sets constructed from the set of finite labelled causal sets. In
the growth process, any $c_n$ is assigned a probability $\mathcal P(c_n)$ and hence we can take
$\mathcal P(\cyl(c_n)) \equiv \mathcal P(c_n)$. This measure is known to extend to all of $\mathcal
S$ by the Kolmogorov-Caratheodary-Hahn extension theorem, so that the triple $(\Omega, \mathcal S, \mathcal P)$
is a well defined probability measure space.  A label invariant measure space can then be
constructed formally \cite{observables} by taking its quotient under relabellings: if
$\widetilde{\Omega}$ is the set of unlabelled past finite causal sets, then $\widetilde{\mathcal S}$
is the sub-sigma algebra of $\mathcal S$ constructed as follows: a subset $\tilde S \subset
\widetilde{\Omega} $ is an element of $\widetilde {\mathcal S}$ if for every $c \in \tilde S$, its
relabelling $c'$ is also in $\tilde S$. The restriction of $\mathcal P$ to $\widetilde{\mathcal S}$
then provides a label invariant measure space $(\widetilde \Omega, \widetilde{\mathcal S}, \mathcal
P_{\widetilde{\mathcal S}})$ \cite{observables}.

But what are these label invariant or covariant events and can they be assigned a physically useful
meaning? Of interest to quantum cosmology are covariant questions of the type: how many bounces has
the universe gone through?  In \cite{observables} it was shown that the formal event algebra
$\widetilde{\mathcal S}$ can be replaced, upto set of measure zero, with a sub-sigma algebra
$\mathcal S'$, which is generated from physically well defined events. A finite unlabelled
sub-causal set $\tilde{c_n} $ of $ \tilde c \in \tilde \Omega $ is said to be a partial stem if it
contains its own past, and is the causal set analogue of a ``past-set'' $J^-(X)$ in a spacetime. A
stem set $\stem(\tilde{c_n})$ is a subset of $\tilde \Omega$ such that every $\tilde c \in
\stem(\tilde{c_n})$ contains the partial stem $\tilde{c_n}$. Thus, for these cosmologies it is
possible to ask, for example, what the probability is for the universe to have a single element
which lies to the past of every other element (originary growth). For transitive percolation, this
is given by the Euler function $\phi(q)= \Pi_{i=1}^\infty (1-q^i)$, where $q=1-p$ is the probability
for a new element to be unrelated to all other elements \cite{djss}.  Thus, the importance of this
sub-sigma algebra is clear -- it makes it possible to pose covariant cosmological questions and to
therefore construct the covariant ``observables'' or events relevant to cosmology. It is the hope
that at least a part of this analysis survives into a quantum dynamics for causal sets.

These stochastic growth models in fact admit a simple quantum generalisation with the transition
probabilities replaced by transition amplitudes. Instead of a
probability measure space the growth process generates a quantum pre-measure space $(\Omega,
\mathcal A, \mu)$, where 
$\mathcal A$ is the  algebra generated from cylinder sets, closed only under finite
unions and intersections, and $\mu$ is a quantum pre-measure $\mu:\mathcal A \rightarrow \mathbb R^+$ which satisfies the
more general quantum sum rule (QSR) 
\cite{qmeasure,rob,qcovers}:
\begin{equation}
\mu(A \cup B \cup C) = \mu(A \cup B) + \mu(A \cup C) + \mu(B\cup C) - \mu(A) -\mu(B) -\mu(C),   
\end{equation} 
for the disjoint sets $A,B, C \in \mathcal A$. $\mu$ is not in general additive since there can be
interference between different sets in $\mathcal A$. This means that the extension of $\mu$ to the
full event sigma algebra $\mathcal S$ is not guaranteed. But such an extension is required  if
we want to construct covariant observables in analogy with the classical growth models.  Using the
Histories Hilbert space construction of \cite{djs} the quantum pre-measure can be recast as a vector
pre-measure \cite{djss} for which a Kolmogorov-Caratheodary-Hanh type extension theorem exists provided $\mu$
satisfies certain condition. This analysis was used to explore certain simple ``complex
percolation'' models in \cite{djss}.  These models are useful quantum generalisations of the
classical stochastic theories since they incorporate both the Markovian evolution as well as label
invariance\footnote{However, is not quite clear what the proper quantum analog of the Bell causality
condition is \cite{joebell,joebelltwo}. While its easy to implement a classical analog of this condition in
these simple models, a more general implementation seems riddled with complications.}.

Underlying the above construction of a quantum measure space is another deep current of enquiry into
the measurement question of quantum theory. The interpretational problems of quantum theory become
of tantamount importance in quantum cosmology where one needs to give meaning to cosmological events
in the absence of measuring devices. A quiet revolution in quantum interpretation was initiated by
Sorkin in a series of papers starting from a { quantum measure} formulation \cite{qmeasure,rob} and
culminating more recently in the anhomomorphic logic proposal \cite{alog}. This is a realistic
interpretation of quantum theory, independent of external observers and measuring devices, which
nevertheless passes the stringent Kochen-Specker test \cite{DG}.  While an exciting development for
quantum theory, a more detailed description is outside the scope of this review. From a functional
point of view however, it  suffices to think of causal set dynamics simply as a quantum measure
space. Questions related to observation reduce to events that are measurable -- an event of quantum
measure zero, for example, can be taken to ``not occur''.

The construction of the causal set action described in Section \ref{kinematics} on the other hand
allows a ``top-down'' approach to quantisation along more conventional lines (keeping the
interpretational issues at bay).  Using a parameter Wick rotation of the type described in
\cite{rafeuc} the Lorentzian action can be Euclideanised to obtain a partition function for causal
sets. Importantly, unlike the standard Euclideanisation which changes the set of histories from
Lorentzian to Euclidean geometries, this procedure does not change the space of causal
sets\footnote{Indeed, causal sets are so inherently Lorentzian it is difficult or impossible to
  imagine a Euclideanised version!}. Causal set dynamics can then be simulated using Monte Carlo
methods and efforts in this direction are currently underway \cite{hrss}. Early simulations show
promise, though much work remains to be done in understanding the complex phase structure of the
entropy landscape.  While an important direction to pursue, however, from a fundamental point of
view it is unclear that the causal set action {\it should} provide us with the correct quantum
dynamics. It is entirely possible that dynamics is dictated by yet undiscovered fundamental rules
and that the Einstein-Hilbert action makes its appearance only as an effective dynamics much like
the Navier-Stokes equation.

An interesting, though limited, example of a top-down approach to constructing causal set quantum is
a 2d model of causal sets which exhibits unexpected results \cite{2dqg}. It is important to
reiterate that unlike most other theories of quantum gravity in which dimension is inbuilt, $\Omega$
includes causal sets that approximate to spacetimes of {\it all} dimensions. Hence much more is
expected from causal set dynamics than from other approaches, since spacetime dimension is also
required to be a prediction of the theory.  On the other hand it is a useful exercise to examine
such a dimensional restriction of causal sets. All finite element causal sets that are approximated
by conformally flat 2-d spacetimes with trivial topology belong to the class of 2d-orders, obtained
by taking the intersection of two linear or total orders. One can think of these two total orders as
the discrete lightcone coordinates $(u,v)$ of a causal set that faithfully embeds into such a
spacetime -- an element $e_1=(u_1,v_1)$ precedes an element $e_2=(u_2, v_2)$ iff both $u_1< u_2$ and
$v_1 < v_2$.  Note that in a sprinkling, the probability of an element to lie exactly on the light
cone of another element (i.e., either their $u$ or $v$ coordinates are the same) is precisely zero.
However, not all 2d orders admit a continuum approximation, and thus, though there is an intrinsic
causal set dimension given by the number of total orders, this does not itself imply
manifoldlikeness. In particular it means that the entropy problem could also potentially rear its
head in such a model. What is surprising is that even for a uniform measure on 2d orders, the
dominant contribution comes from those  which faithfully embed into {\it flat} 2d spacetime
\cite{2dqg, ES,winkler}. A uniform measure can be motivated from a continuum inspired model of
quantum gravity \cite{2dqg} which soups up this order theoretic result into a genuine 2d model of
quantum gravity.  Generalisations to higher dimensions though much harder to do, would be a natural
direction to pursue.

\section{Conclusions} \label{conclusions} 

What is the current state of causal set theory? For years, criticisms have ranged from saying, on
the one hand, that the approach is too minimalist to give us any recognisable physics, and on the
other that it is too ad hoc and not well motivated enough. The latter complaint in my opinion is
entirely unjustified. {\it All} approaches to quantum gravity necessarily use new ingredients, even
when guised in the modest robes of ``naturalness''.  Questions of choice and emphasis are motivated
either by principles of simplicity (however that is interpreted) and by phenomenological
restrictions. Causal set theory is well motivated in this sense -- a causal set is the simplest
possible manifestation of a quantum {\it Lorentzian} geometry. It takes the Lorentzian character of
spacetime, in particular the {\it classical} results of \cite{HKM,Malament}, most seriously adding
only the simplest ingredient of discreteness in a manner {\it consistent} with the classical
results. They are also, at a basic level, phenomenologically justified since they satisfy Lorentz
invariance, a well verified symmetry of nature.  The fact that it is too minimalist is a criticism
that perhaps understandably kept interested researchers at bay. It was unclear for a long time how
to recover locality from such a fundamentally non-locality theory, without going to the continuum
limit.  In retrospect, these fears are unfounded given the recent developments in the field, in
particular the construction of a well defined local D'Alembertian operator and Einstein-Hilbert
action for causal sets. It gives hope that with continued work other seemingly insurmountable
hurdles could also be one day be cleared.
 
I now present a short list of questions of immediate interest to causal set workers. It is far from
exhaustive, but hopefully it is representative. Arguably, one of the most interesting questions is
whether the formalism developed in \cite{stevenone,steventwo} can be applied to black-holes to study
Hawking radiation. Discreteness is often taken to be synonymous with a ultraviolet cut-off, but this
is not true in causal set theory, since the discretisation is Lorentz invariant. It is therefore not
clear whether causal sets cure the transplankian problem. Is there instead a manifestation of
non-locality which modifies the thermal spectrum as in other non-local theories
\cite{noncommutative}? Constructing a decoherence functional (or quantum measure) in terms of the
Feynman propagator is also an important question to address, as is the question of how interactions
manifest themselves in this framework. In \cite{steventhesis} an interesting proposal has been made
for constructing an S matrix on the causal set for interacting fields and it is important to know
how far this analysis can be taken.  In \cite{BD,Glaser,locality} in the process of constructing
discrete operators, non-local continuum actions were constructed and it would be an interesting
exercise to study such non-local field theories and obtain constraints on the non-locality scale.

The numerical work on the Monte Carlo simulations for quantum dynamics of causal sets, while still
in its infancy, should be able to tell us whether a Euclideanised dynamics of causal sets can give
rise (after a reverse Wick rotation) to causal sets that are manifoldlike in the classical limit.
In the Causal Dynamical Triangulation(CDT) approach, a Euclideanised dynamics has been employed
successfully to cure the entropy problem encountered when the causality constraint is not included
\cite{CDT} and it is a somewhat analogous success that we seek with causal sets. On the other hand,
while interesting and of value, the success or failure of this approach may not be related, in a
simple way, to the construction of a truly bottom-up quantum dynamics. 

As described in Section \ref{dynamics} one would like to obtain a fundamental class of quantum
dynamics based on first principles much like the classical stochastic growth dynamics, but the
implementation of causality remains an important open question \cite{joebell,joebelltwo}.  In
addition, any examination of quantum dynamics ties in intimately with the measurement problem (which
we have touched upon perhaps too lightly in this review.) The quantum measure approach provides a
framework to do so and while one now has an understanding of certain finite dimensional situations,
developments in the infinite dimensional case will be crucial to quantum gravity \cite{djss}.

Another very important question which, when answered, should make the approach more amenable to the
wider community, is how matter should be incorporated into causal set theory. The combination of
discreteness with Lorentz invariance means that there is no ready prescription for putting vector
fields or higher spin fields on the causal set. Should these instead emerge from the theory or does
one simply need an (as yet undiscovered) well defined way of putting them in ``by hand''? Fermions
are perhaps easier to incorporate -- one can assign spins to the links, but relating these to
spacetime fermions brings us back to the problem encountered with putting higher spins on the causal
set. Progress on such questions would be extremely valuable in constructing more realistic
phenomenological models. In \cite{steventhesis} a suggestion for a Feynman chequerboard model was
made for introducing spin on the causal set. These discussions suggest that rather than getting
classical fields to sit on a causal set, one should construct the more fundamental quantum field
operators which in the continuum approximation will give rise to the classical fields.

Another very important direction that needs to be actively pursued is the search for
phenomenological signatures of the sort described in Section \ref{phen}. To begin with, more
realistic models of a varying $\Lambda$ need to be constructed to match observations more precisely.
Can the neutrino origins of cosmic rays give hope to the swerve model of particles and are there
other ways in which the Lorentz invariant discreteness of causal sets can manifest itself in
observations?  One can at this stage ask bolder questions -- can causal set theory explain the other
big puzzles of cosmology like dark matter or structure formation, and can it provide an alternative
explanations to the scale invariance of the CMB power spectrum? These may seem too broad a class of
questions, but it seems possible at this point in the development of the theory, to begin building
more realistic models to constrain the effects of spacetime discreteness on observations.

In this article I have outlined causal set theory in broad strokes and pointed to the directions
which are currently being pursued. A modest hope is that the reader will become familiar with the
basic tenets of this approach, while a more ambitious hope is that she or he will then begin
actively ruminating on causal sets!  

\vskip1cm

\noindent {\bf Acknowledgements:} I would like to thank my causal set co-workers, especially Rafael
Sorkin, Fay Dowker, David Rideout and Joe Henson for valuable discussions over the years. This work
was supported in part by the Royal Society International Joint Project 2006-R2.

\end{document}